\documentclass[twocolumn]{article}

\usepackage[american]{babel}

\usepackage{booktabs}
\usepackage{graphicx} 
\usepackage{amsfonts}
\usepackage{amsmath}
\usepackage{xcolor}
\usepackage{subcaption}

\title{Quantum-Assisted Generative Anomaly Detection: A Dimension-Controlled Study of Latent Refinement}
\date{November 2026}

\author{
Thesath Wijayasiri\textsuperscript{1},
Omar Bacarreza\textsuperscript{2},
William Clements\textsuperscript{2},\\
Kar Wai Fok\textsuperscript{1},
Vrizlynn L. L. Thing\textsuperscript{1}\\[0.5em]
\small \textsuperscript{1}ST Engineering\\
\small \textsuperscript{2}ORCA Computing
}

\begin{document}

\maketitle
\begin{abstract}
Anomaly-based malware detectors learn what benign files look like and flag deviations. Generative approaches to this problem can make use of a ``latent" or ``prior" data representation, which is often chosen to be a Gaussian distribution. We ask whether the choice of prior distribution matters, and whether one produced by quantum hardware behaves differently from a classical one. Building on SAGEGAN, a style-based detector trained only on benign data, we compare three latent distributions: the original 512-dimensional Gaussian, a 48-dimensional Gaussian, and a 48-dimensional distribution sampled from an ORCA Computing PT-2 photonic quantum processor. We then refine each trained model's frozen representation using greedy and reinforcement-learning methods. Across ten random seeds, the quantum-derived model shows richer internal structure with better effective rank and latent similarity. It also benefits most from refinement. Though final detection accuracy is similar to that of the refined Gaussian latent, these findings show differences in latent geometry and refinement behaviour between the trained configurations.
%This study investigates how a photonic quantum derived latent prior affects the learned representation and post training refinement of a generative anomaly detection model. Using SAGEGAN as the underlying framework for malware detection, we compare a 512 dimensional Gaussian prior, a 48 dimensional Gaussian prior, and a 48 dimensional PT prior sampled from an ORCA Computing PT-2 photonic quantum processor. The 48 dimensional Gaussian configuration provides a dimension matched control for assessing the effect of the latent distribution. Greedy and reinforcement learning based refinement are applied to the encoded style representations after training, while the anomaly detector remains frozen. Evaluation across ten random seeds considers test AUC, refinement gains, effective rank, and latent prior matching. The PT model consistently shows higher effective rank and closer latent prior matching than the Gaussian configurations. It also achieves the largest mean improvement from refinement, with mean test AUC increasing from 0.8570 to 0.8935 following reinforcement learning based refinement. The corresponding refined AUC for the dimension matched Gaussian model is 0.8905.However, despite the difference in final AUC not being statistically significantm, these findings show differences in latent structure and refinement behaviour.
\end{abstract}
\section{Introduction}

Malware detection models need to identify malicious files that may differ from the examples available during training. Generative anomaly detection approaches learn a model of benign data and use deviations from that model
to score new inputs \cite{Schlegl2017AnoGAN}. In a GAN-based detector, the latent prior is the distribution from which inputs to the generator are sampled during training. Its suitability depends on the target data and
the capacity of the generator \cite{Hu2023ComplexityLatent}. Photonic measurements provide a source of non-Gaussian samples whose coordinates can exhibit statistical dependencies \cite{Bacarreza2026QuantumLatent}.
We ask whether training with such samples changes the representation learned from benign executables, detection performance, and the response to post-training refinement.

A style-based GAN maps latent samples to style vectors that modulate successive stages of image synthesis \cite{Karras2020StyleGAN2}. This
allows Gaussian and photonic-derived samples to feed the same generator design through its style mappings. The layer-wise structure
also gives refinement a way to adjust the style supplied to an individual synthesis stage.

We use SAGEGAN because it combines this generator design with an encoder trained on benign executable images \cite{Wijayasiri2026SAGEGAN}. The encoder maps each input directly to seven layer-specific style vectors
that the generator uses to reconstruct the image. Prior alignment and latent consistency connect the sampling and encoding pathways during
training. The encoded styles can therefore be examined directly, used to reconstruct the input for anomaly scoring, and refined after training with the detector frozen.

We compare a 512-dimensional Gaussian reference, a 48-dimensional Gaussian configuration, and a 48-dimensional configuration using
precollected measurements from an ORCA Computing PT-2 photonic quantum processor. 
The photonic prior width is set by the $48$ measured output qumodes
of the PT-2. The $48$-dimensional Gaussian control matches this width
so the two configurations can be compared at the same latent and
per-layer style dimensionality.
The two 48-dimensional configurations use the same detector architecture and $7\times48$ style tensors. We train each configuration independently and compare base detection performance with performance after greedy and reinforcement-learning refinement of the encoded styles.

We evaluate model performance using detection performance, effective rank, and latent matching. Effective rank provides an indication of how broadly variation is distributed across the latent dimensions, while latent matching assesses the alignment between encoded benign representations and the corresponding generative prior. These measures allow us to examine differences in the latent representation alongside changes in detection AUC.

Across the evaluated runs, the model with the quantum latent consistently shows higher effective rank and better latent matching than the Gaussian configurations. It also achieves the largest mean improvement from latent refinement. While the refined quantum and Gaussian latents reach similar final detection accuracy, these results show that the PT configuration exhibits higher normalized effective rank and closer moment matching, suggesting that its structural properties merit further investigation.

\begin{figure*}[t]
    \centering
    \includegraphics[width=\textwidth]{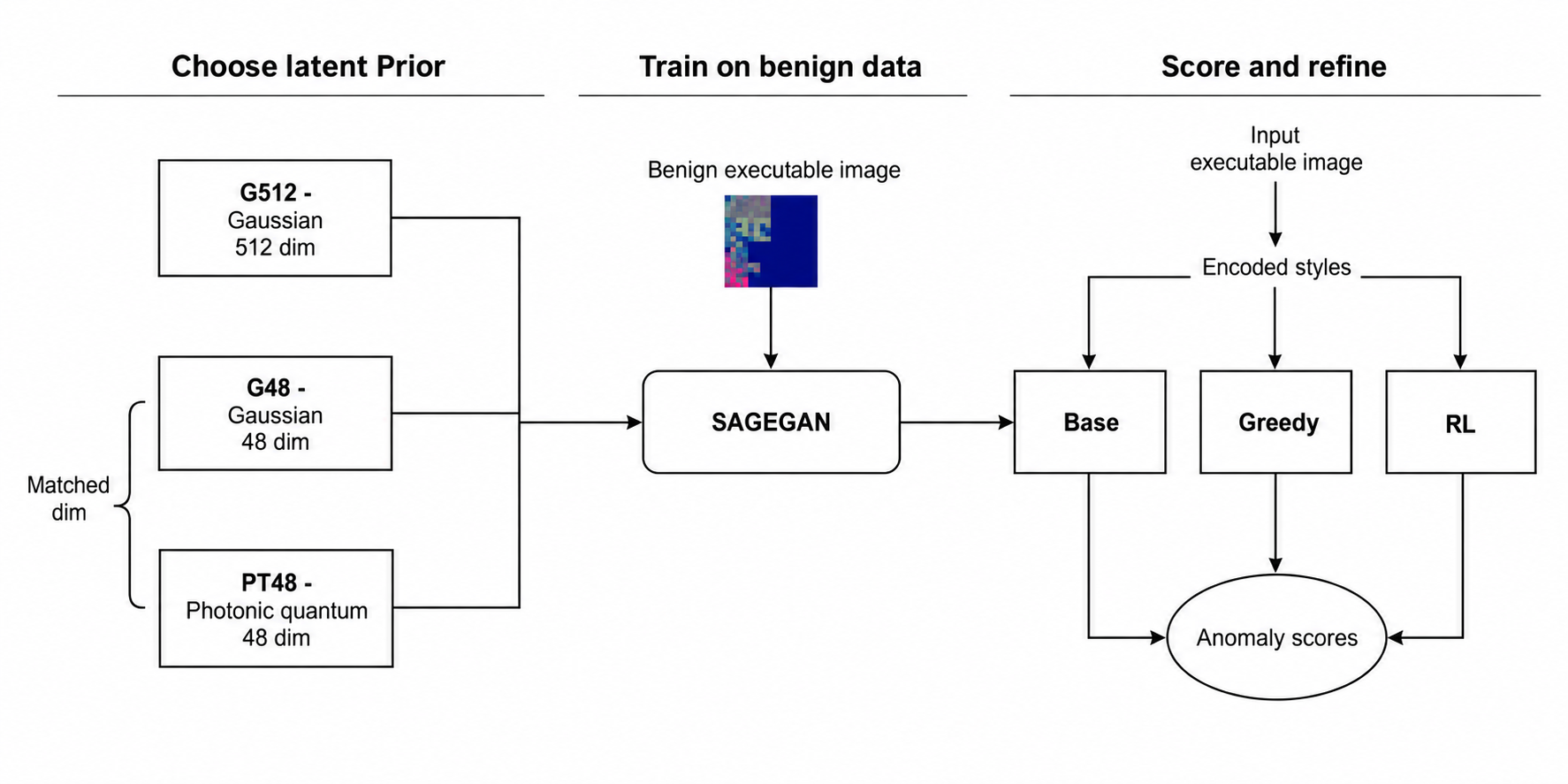}
    \caption{Conceptual overview of the study.
    Separate SAGEGAN detectors are trained on benign
    executable images using Gaussian or photonic-derived priors. Each trained detector is then frozen
    and evaluated without refinement, with greedy
    refinement, and with RL refinement. Detection
    performance, latent diagnostics, and refinement
    gains are compared across ten seeds.}
    \label{fig:design}
\end{figure*}
 
Figure~\ref{fig:design} summarises the study, from latent
prior selection and benign-only detector training to
evaluation with and without post-training refinement.

\section{Related Work}
\label{sec:related_work}

\subsection{Generative Anomaly Detection}

One-class anomaly detection learns a representation of normal data and identifies samples that deviate from it. Approaches such as Deep Support Vector Data Description (SVDD) learn this representation by mapping normal samples into a compact region of feature space using an objective designed for anomaly detection \cite{Ruff2018DeepSVDD}. Generative approaches instead learn to reconstruct or generate normal samples and use departures from the learned distribution to assess abnormality.

Generative adversarial network (GAN) methods provide one approach to modelling the data distribution. AnoGAN searches for a latent representation of each input through iterative optimisation and combines reconstruction and discriminator feature differences to obtain an anomaly score \cite{Schlegl2017AnoGAN}. The subsequent f-AnoGAN framework uses a trained encoder to map inputs into the latent space, enabling faster inference while retaining reconstruction and discriminator feature errors for scoring \cite{Schlegl2019fAnoGAN}.

Other approaches incorporate consistency between the input and its latent representation. GANomaly uses an encoder--decoder--encoder architecture and learns from both image reconstruction and latent discrepancies \cite{Akcay2019GANomaly}. Adversarially Learned Anomaly Detection (ALAD) uses bidirectional adversarial learning with consistency constraints in both data and latent spaces \cite{Zenati2018ALAD}. These methods provide relevant foundations for examining the relationship between reconstruction, latent consistency, and anomaly detection.

For malware analysis, Nataraj et al.\ represented executable binaries as images and used visual features for malware classification \cite{Nataraj2011MalwareImages}. The SAGEGAN framework applies style-based generative modelling to executable images and learns from benign samples for anomaly detection \cite{Wijayasiri2026SAGEGAN}. The present work retains this backbone and image representation while examining prior choice and refinement.

\subsection{Prior Design in Generative Models}

The prior determines the distribution of inputs mapped through a generator. Its suitability depends on both the target data and the capacity of the model. Hu et al.\ study this relationship through generator complexity and characterise a data-dependent latent distribution that makes effective use of the generator's capacity \cite{Hu2023ComplexityLatent}. Their work motivates considering the latent distribution as part of model design.

Several methods have been introduced for learning a structured prior. Tomczak and Welling introduce VampPrior, a mixture of variational posteriors conditioned on learnable pseudo-inputs \cite{Tomczak2018VampPrior}. Although developed for variational autoencoders, it provides an example of how a flexible prior can improve generative modelling beyond a simple fixed distribution.

Style-based architectures introduce a further distinction between sampled latent inputs and the style representations used to condition the generator. StyleGAN2 uses such representations within its generator architecture \cite{Karras2020StyleGAN2}. Prior choice therefore needs to be considered alongside the transformation into style space. For the present study, this motivates comparing Gaussian and photonic-derived priors at the same latent and style dimensionality.

\subsection{Quantum-Derived Latent Distributions}

A quantum processor can serve as the source of latent samples for a classical generative network. Wilson et al. modelled the latent distribution of an associative adversarial network with a quantum annealer \cite{wilson2021quantum}. Rudolph et al. sampled a GAN prior from a quantum circuit on an ion-trap processor to generate handwritten digits \cite{rudolph2022generation}. Related approaches have been applied to molecular generation \cite{kao2023exploring,ghazi2025quantum} and image reconstruction \cite{xiao2024quantum}. Jin and Merz replaced or combined Gaussian inputs to BigGAN with quantum-correlated samples collected before training, and reported improved generation \cite{jin2025improving}.

Bacarreza et al. showed that, under assumptions on generator capacity, a latent distribution sampleable only by a quantum device yields pushforward distributions that no efficiently sampleable classical latent can approximate \cite{Bacarreza2026QuantumLatent}. On synthetic and molecular datasets, quantum samples outperformed several classical priors. The same approach has since supported experimentally validated peptide design \cite{engdal2026hybrid}. Bacarreza et al. also reported negative results for StyleGAN on CIFAR-10, indicating that the benefit depends on architecture and data, which is relevant to the style-based detector studied here.

The closest precedent for anomaly detection is Kailasanathan et al. \cite{kailasanathan2026quantum}. They trained an ensemble of GAN discriminators on normal operating data from a simulated biomanufacturing process, using latents from a Gaussian distribution and a quantum processor. Both quantum latents improved detection over the Gaussian baseline. The present study extends this line of work to a style-based encoder–generator detector with reconstruction- and refinement-based scoring.

The latter three studies (\cite{Bacarreza2026QuantumLatent,engdal2026hybrid,kailasanathan2026quantum}) all used photonic quantum processors, including the ORCA Computing PT-2. Several properties motivate this choice. First, sampling from such a device is believed to be classically intractable beyond a few tens of photons \cite{aaronson2011computational}. Second, its outputs are highly non-uniform \cite{aaronson2013bosonsampling}. Third, multi-photon interference produces correlations at multiple orders \cite{Bacarreza2026QuantumLatent}, creating a rich statistical structure distinguishing them from simple classical priors. Moreover, loss, which is the main error mechanism of photonic systems, does not drive outputs toward uniformity in the way decoherence does in other systems. Decoherence tends to drive circuit outputs toward the uniform distribution \cite{wang2021noise}, erasing the statistical structure we aim to leverage with a structured latent distribution. Kailasanathan et al. additionally noted the practical accessibility of room-temperature, commercially available photonic hardware \cite{kailasanathan2026quantum}. For the same reasons, we use samples from a PT-2 processor in this work.

\subsection{GAN Inversion and Post-Training Latent Refinement}

GAN inversion seeks a latent representation that allows a trained generator to reconstruct a given input. Creswell and Bharath investigate inversion into the latent space of a pretrained GAN and use reconstruction to examine the information captured by the model \cite{Creswell2018GANInversion}. AnoGAN applies related latent optimisation principles to anomaly detection \cite{Schlegl2017AnoGAN}.

Learned inversion methods also support iterative updates. ReStyle predicts residual changes to a StyleGAN latent representation, progressively improving the reconstruction over successive encoder passes \cite{Alaluf2021ReStyle}. This provides a close connection to refinement after an initial encoding, although its update mechanism and evaluation objective differ from those considered here.

In our framework, greedy search and a learned policy operate on encoded style tensors while the anomaly detector remains frozen. The policy uses REINFORCE, following the policy-gradient formulation introduced by Williams \cite{Williams1992REINFORCE}. We evaluate how the resulting changes contribute to anomaly scoring, since improving reconstruction alone does not establish improved detection.

\section{Methodology}
\label{sec:methodology}

The present study brings these research directions together through a comparison of 512-dimensional Gaussian, 48-dimensional Gaussian, and 48-dimensional photonic-derived priors within SAGEGAN. The dimension-matched comparison examines prior choice, while the greedy and learned refinement conditions assess how the resulting representations respond to further optimisation. Detection results and latent diagnostics are considered together to examine prior alignment, representation geometry, and refinement behaviour.

\subsection{Overview and Experimental Design}
\label{sec:overview}

Generative anomaly detectors identify malicious files by learning a model of
benign data and flagging inputs that this model represents or reconstructs
poorly. In style-based detectors such as SAGEGAN \cite{Wijayasiri2026SAGEGAN},
the benign model is organised around a \emph{prior distribution}: a distribution from
which style representations are drawn and to which encoded benign inputs are
aligned. The distribution is therefore a critical part
of the representation the detector learns. This motivates the two questions we
address. First, does the structure of the prior affect anomaly detection
performance? Second, do different priors respond differently when their learned
style space is refined after training?

To examine these questions, we compare prior configurations within the SAGEGAN framework.
The anomaly detection task,
the binary-to-image representation, and the encoder, generator, and discriminator
architectures are all inherited unchanged from \cite{Wijayasiri2026SAGEGAN}
(Section~\ref{sec:backbone}). Against this fixed backbone we study two axes of
variation: the choice of prior, and the application of post-training
latent refinement.

The first axis concerns the prior itself. Our primary interest is a
$48$-dimensional prior sampled from an ORCA Computing PT-2 photonic quantum
processor, which we compare against the original $512$-dimensional Gaussian prior
of SAGEGAN. A direct comparison between these two, however, would confound two
distinct changes: the change in distribution, from Gaussian to
photonic-derived, and the change in dimensionality, from $512$ to $48$.
To isolate the effect of the latter, we introduce a third configuration: a
dimension-matched $48$-dimensional Gaussian prior that shares the PT
configuration's latent and style dimensionality but retains a classical Gaussian
distribution. This yields three configurations, denoted $G512$, $G48$, and
$PT48$. In every case a base latent vector is transformed into seven
layer-specific style vectors, one for each modulation stage of the generator,
giving the style tensors
\begin{equation}
\begin{aligned}
W_{G512} &\in \mathbb{R}^{7\times512},\\
W_{G48}  &\in \mathbb{R}^{7\times48},\\
W_{PT48} &\in \mathbb{R}^{7\times48}.
\end{aligned}
\end{equation}
The three configurations support two controlled comparisons. Contrasting $G512$
with $G48$ holds the distribution fixed and isolates the effect of
dimensionality, while contrasting $G48$ with $PT48$ holds the dimensionality
fixed and compares the two prior configurations. 

The second axis concerns what happens after training. In a first stage, the
three configurations are trained independently as complete SAGEGAN detectors. In
a second stage, the trained detector (the encoder, generator, and
discriminator) is frozen, and only the encoded style representation assigned
to an input is refined. This refinement proceeds as a short sequence of small discrete steps that move it
toward the benign manifold, and the number of steps required to perform this move provides a useful anomaly detection signal. Separating training from refinement
distinguishes a representation's initial quality from its \emph{navigability}: the
extent to which its style space can be traversed toward a more representative
benign region once the detector is fixed. We apply two refinement methods to each
trained configuration: a deterministic greedy search and a learned
reinforcement-learning (RL) policy. The greedy method acts as a non-learned
control, so that the sequence
\begin{equation}
\text{Base}\rightarrow\text{Greedy}\rightarrow\text{RL}
\end{equation}
distinguishes improvements that follow from the availability of a useful local
search space from any additional benefit contributed by a learned policy.

Taken together, these two axes examine three aspects: the effect of latent dimensionality ($G512\rightarrow G48$), differences between prior configurations at matched dimensionality
($G48\rightarrow PT48$), and the responsiveness of each learned representation to
post-training optimisation ($\text{Base}\rightarrow\text{Greedy}\rightarrow
\text{RL}$). The design allows a configuration's base detection performance and
its refinement responsiveness to be analysed as distinct properties of the
learned representation.

Figure~\ref{fig:design} summarises the overall design. The remainder of this
section describes the fixed backbone (Section~\ref{sec:backbone}), the three
priors (Section~\ref{sec:priors}), encoder training and base scoring
(Section~\ref{sec:encoder}), and the refinement procedure
(Section~\ref{sec:refinement}).

\subsection{SAGEGAN Backbone}
\label{sec:backbone}

All three configurations share a single SAGEGAN detector, inherited unchanged
from \cite{Wijayasiri2026SAGEGAN}. This section describes that fixed substrate.
Everything in it is held constant across $G512$, $G48$, and $PT48$; the only
quantity that varies is the latent dimensionality $d$, and the way it enters is
made explicit at the end of the section. 
We selected the detector architecture to support both a comparison of training priors and refinement at individual synthesis stages. In
f-AnoGAN and ALAD, the encoder produces a single latent code at the generator input, so refinement would act on that shared code
\cite{Schlegl2019fAnoGAN,Zenati2018ALAD}. StyleGAN2 provides layer-wise synthesis controls and uses latent optimization to invert an input image \cite{Karras2020StyleGAN2}. SAGEGAN combines layer-wise synthesis with an encoder trained on benign executable images that directly estimates one style vector per stage, alongside a separate pathway for samples drawn from the training prior \cite{Wijayasiri2026SAGEGAN}. This lets us compare Gaussian and photonic priors at matched dimensionality and apply the same refinement rules to the resulting encodings, including moves that target an individual synthesis stage.

\subsubsection{Image Representation and Architecture}

Each executable is converted into a $256\times256$ three-channel image using the
preprocessing procedure of \cite{Wijayasiri2026SAGEGAN}. The detector comprises an encoder $E$, a style-conditioned generator $G$, and a
discriminator $D$. The generator reconstructs an input from a learned $4\times4$
constant through seven style-modulated convolutional stages, each conditioned on
its own style vector. The discriminator provides adversarial supervision during
training and subsequently contributes a realism term to the anomaly score
(Section~\ref{sec:encoder}).

The detector admits two distinct pathways into these style-modulated stages,
which it is useful to separate. In the \emph{analysis} pathway, used at inference,
the encoder maps an input image directly to a style tensor
\begin{equation}
W_E = E(x) = \{w_1, w_2, \ldots, w_7\}, \qquad w_l \in \mathbb{R}^{d},
\end{equation}
which the generator then decodes to a reconstruction $G(W_E)$. In the
\emph{sampling} pathway, used during training, the style vectors are instead
produced from a latent sample drawn from the prior, as described below. The
encoder, generator, and discriminator are agnostic to which prior is in use.

\subsubsection{Layer-Wise Style Construction}
\label{sec:style}

In the sampling pathway, a base latent vector $z\in\mathbb{R}^{d}$ drawn from the
selected prior is transformed independently for each generator stage. Seven
learnable affine mappings
\begin{equation}
A_l:\mathbb{R}^{d}\rightarrow\mathbb{R}^{d}, \qquad l=1,\ldots,7,
\end{equation}
produce one style vector per stage, $w_l = A_l(z)$, so that the generator receives
\begin{equation}
W(z) = [A_1(z), A_2(z), \ldots, A_7(z)] \in \mathbb{R}^{7\times d}.
\label{eq:style-tensor}
\end{equation}

The dimensionality $d$ is the only difference in model architecture between configurations: $d=512$ for $G512$ and $d=48$ for both $G48$ and $PT48$.

\subsection{Prior Configurations}
\label{sec:priors}

The three configurations differ in the base distribution $P(z)$ from which
the latent sample $z\in\mathbb{R}^{d}$ is drawn in the sampling pathway of
Section~\ref{sec:style}. In every case the sample is passed through the same
seven-stage affine construction of Equation~\ref{eq:style-tensor} and the same
generator conditioning. 

\subsubsection{Gaussian Priors: Reference and Dimension-Matched Control}
\label{sec:gaussian}

Two of the three configurations use the learnable diagonal Gaussian prior of the
original SAGEGAN \cite{Wijayasiri2026SAGEGAN}, differing only in dimensionality $d=48$ and $d=512$.
A base sample is drawn as
\begin{equation}
z_G = \mu + \sigma \odot \epsilon, \qquad \epsilon \sim \mathcal{N}(0, I_d),
\end{equation}
with learned $\mu,\sigma\in\mathbb{R}^{d}$, and is then contracted toward the
learned mean by a truncation coefficient $\psi$ set to 0.85,
\begin{equation}
z_G^{(\psi)} = \mu + \psi\left(z_G - \mu\right),
\end{equation}
before the affine mappings of Section~\ref{sec:style} produce the style tensor.

\subsubsection{Photonic Quantum-Derived Prior}
\label{sec:pt48}

The focus of our work involved replacing Gaussian sampling with a $d=48$ dimensional
prior obtained from an ORCA Computing PT-2 photonic
quantum processor.
Quantum-generated latents have been investigated as alternatives to conventional
continuous priors in generative models
\cite{rudolph2022generation,jin2025improving,Bacarreza2026QuantumLatent,engdal2026hybrid},
with \cite{Bacarreza2026QuantumLatent} hypothesising that such distributions can
be better suited than common classical priors to generating complex, multimodal
data.

The PT-2 implements the boson sampling model of computation: $N$ single photons
are injected into an $M\times M$ linear interferometer ($M\geq N$), producing a
large entangled state, and a measurement returns a length-$M$ vector recording how
many photons are detected in each output channel, or ``qumode''. Because the
measurement is probabilistic, each run typically yields a different photon
pattern, and the distribution over outcomes is intractable to sample classically beyond a few tens of photons
\cite{aaronson2011computational}. Although boson sampling is not universal for
quantum computation, \cite{Bacarreza2026QuantumLatent} argued that its highly
non-uniform output distribution makes it particularly suitable as a prior distribution.

% \begin{figure*}[t]
%     \centering
%     \includegraphics[width=\textwidth]{tbi.png}
%     \caption{Schematic of the two-loop time-bin interferometer
%     (TBI).}
%     \label{fig:tbi}
% \end{figure*}

\begin{figure*}[t]
    \centering
    \begin{subfigure}{0.8\textwidth}
        \centering
        \includegraphics[width=0.9\linewidth]{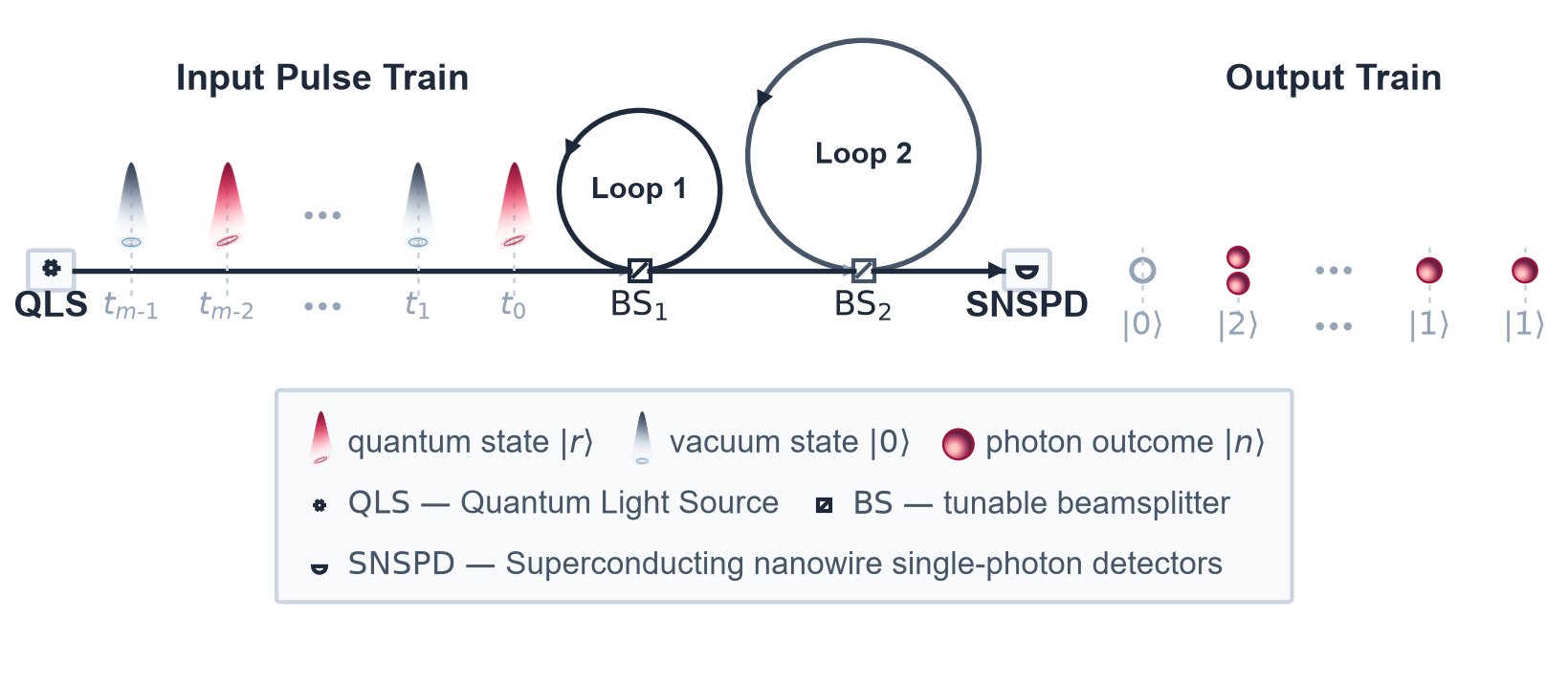}
        \label{fig:a}
    \end{subfigure}

    \begin{subfigure}{0.8\textwidth}
        \centering
        \includegraphics[width=0.9\linewidth]{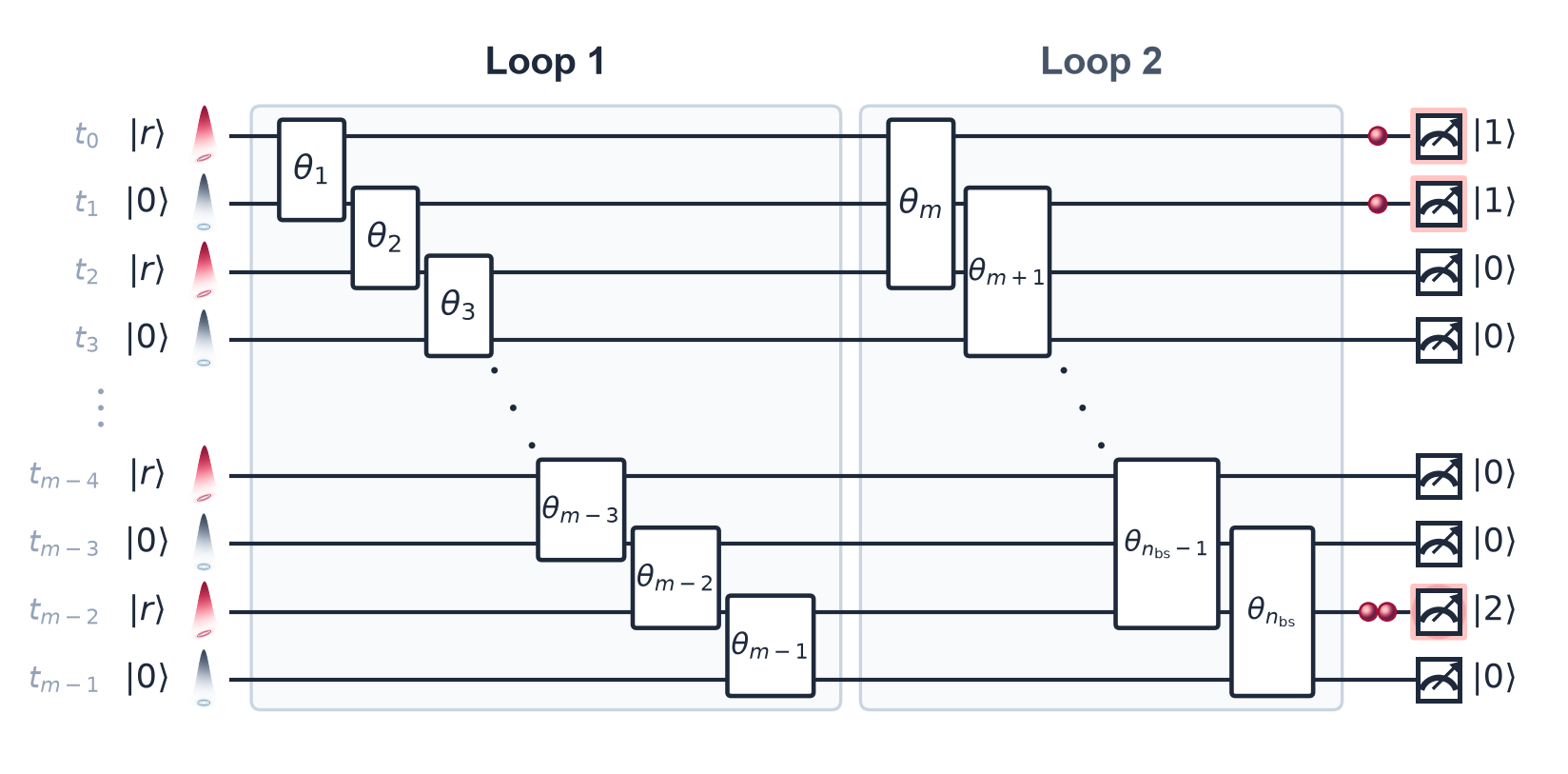}
        \label{fig:b}
    \end{subfigure}

    \caption{Schematic of the photonic quantum processor used in this work.
    Top~Physical diagram of the ORCA Computing PT-2 architecture. A quantum light
    source (QLS) emits a train of $m$ time bins $t_0, \dots, t_{m-1}$. The pulse train passes through two fibre delay
    loops. Each loop is coupled to the main path by a tunable beamsplitter
    ($\mathrm{BS}_1$, $\mathrm{BS}_2$), the reflectivity of which is set
    independently for each time bin. 
    Superconducting nanowire single-photon detectors (SNSPDs) then register
    the outputs, giving a photon-count outcome $\lvert n \rangle$ for
    each temporal mode (qumode).
    Bottom~Equivalent circuit representation. Each time bin corresponds to a
    spatial mode, and each beamsplitter interaction corresponds to a
    two-qumode gate parametrised by an angle $\theta_i$. Loop~1 implements the
    nearest-neighbour gates $\theta_1, \dots, \theta_{m-1}$. Loop~2
    implements the next-nearest-neighbour gates
    $\theta_m, \dots, \theta_{n_{\mathrm{bs}}}$.
    Measuring all $m$ output modes gives one photon-count vector, shown here
    as an example outcome. In this work, $m = 48$, and all $\theta_i$ were
    set to random values that were held fixed throughout collection of the
    one-million-sample pool used as the $PT48$ prior.}
    \label{fig:main}
\end{figure*}

We operated the PT-2 remotely with its programmable interferometer parameters set to random
values, held fixed throughout data collection, to produce a pool of one million
samples across $48$ output qumodes with an average photon number of 17.
Figure~\ref{fig:main} illustrates the time-bin interferometer
(TBI), showing the programmable mixing of temporal modes
and an example photon-count measurement outcome.

The PT-2 setup used in this study provided $48$ output qumodes.
Each measurement therefore yielded a vector with one photon count
per mode. We use the standardized vectors at their native width,
so each measured mode contributes one coordinate to the
$d=48$ base latent before the layer-wise style mappings. Let
\begin{equation}
z_{PT}\sim P_{PT}, \qquad z_{PT}\in\mathbb{R}^{48},
\end{equation}
where $P_{PT}$ is the empirical distribution of the saved sample pool, drawn from
uniformly with replacement. The pool is standardised before use by a single global
mean and standard deviation computed over all entries. This removes simple distributional mismatches as a potential confounding factor between the Gaussian and quantum distributions. The standardised
sample is passed to the same seven affine mappings,
\begin{equation}
W_{PT48} = [A_1(z_{PT}),\ldots,A_7(z_{PT})] \in \mathbb{R}^{7\times48}.
\end{equation}
The comparison between $PT48$ and $G48$ therefore preserves the number of style
stages, the style dimensionality, and the entire downstream generator
conditioning.

\subsection{Encoder Training and Base Anomaly Score}
\label{sec:encoder}

The detector is trained on benign inputs only. For anomaly scoring to work, the
encoder's analysis pathway (Section~\ref{sec:backbone}) must place benign inputs
in the same style region that the generator has learned to decode. Training
therefore pursues two goals at once: benign inputs must reconstruct well, and
their encoded styles must be aligned with the prior. We describe the objectives
that enforce these goals, then the base anomaly score they support. The
adversarial and discriminator objectives are inherited unchanged from
\cite{Wijayasiri2026SAGEGAN} and are not restated here.

\subsubsection{Prior Alignment}

Alignment is enforced through the first two moments of the encoded benign
styles. Let $\mu_E$ and $\sigma_E^2$ denote the empirical mean and variance,
computed across the benign minibatch and the seven style layers, with style
dimension $d$.

For the Gaussian configurations, the reference moments are the learnable
base-prior parameters, $\mu_G=\mu$ and $\sigma_G^2=\sigma^2$, before
truncation and the affine mappings. The alignment objective is
\begin{equation}
\begin{aligned}
\mathcal{L}_{\mathrm{prior}}^{G}
&= \frac{1}{d}\left\|\mu_E-\mu_G\right\|_2^2 \\
&\quad + \frac{1}{d}\left\|\sigma_E^2-\sigma_G^2\right\|_2^2 .
\end{aligned}
\end{equation}
This loss updates the encoder and the Gaussian base-prior parameters, with
$\log\sigma$ used to parameterize the standard deviation.

For the PT configuration, reference moments are estimated from samples drawn
from the fixed measurement pool and passed through the current affine
mappings. Denote these sampled-style moments by $\mu_{PT}^{(s)}$ and
$(\sigma_{PT}^{(s)})^2$. The alignment objective is
\begin{equation}
\begin{aligned}
\mathcal{L}_{\mathrm{prior}}^{PT}
&= \frac{1}{d}\left\|\mu_E-\operatorname{sg}(\mu_{PT}^{(s)})\right\|_2^2 \\
&\quad + \frac{1}{d}\left\|\sigma_E^2-
\operatorname{sg}((\sigma_{PT}^{(s)})^2)\right\|_2^2 ,
\end{aligned}
\end{equation}
where $\operatorname{sg}$ denotes stop-gradient. This loss updates only the
encoder. The PT reference moments are recomputed during training but treated
as constants within each alignment update. Both configurations train their
sampling-path affine mappings through the generator's adversarial objective;
these mappings receive no gradients from the alignment losses above.

Our aim is to obtain the strongest practical version of each model within
the available hardware setting. We therefore preserve SAGEGAN's original
learnable Gaussian prior \cite{Wijayasiri2026SAGEGAN}, while adapting PT
alignment to the fixed measurement pool. This follows benchmarking practice
of using training configurations appropriate to each model
\cite{Bowles2024Benchmarking,RecioArmengol2025Generative}.
Both configurations use the same refinement and evaluation protocol.
The results compare the complete trained configurations, with alignment
strategy remaining a possible contributor to the observed differences.

\subsubsection{Latent Consistency}

A latent-consistency objective is retained across all configurations to encourage
the encoder to invert the generator. A style tensor $W$ is sampled from the
prior, decoded, and re-encoded, and the recovered styles are compared with the
original:
\begin{equation}
\mathcal{L}_{\mathrm{lat}}
= \mathbb{E}_{W\sim p(W)}
\left[ \left\| E(G(W)) - W \right\|_2^2 \right].
\end{equation}
The generated image $G(W)$ is detached when computing this loss, so that the
objective moves the encoder toward the generator's style representation without
reshaping the generated target itself.

\subsubsection{Combined Objective}

The encoder and generator are trained under a reconstruction and prior-alignment
objective,
\begin{equation}
\mathcal{L}_{E,G}
= \lambda_{\mathrm{rec}} \left\| x - G(E(x)) \right\|_1
+ \lambda_{\mathrm{prior}} \mathcal{L}_{\mathrm{prior}} ,
\end{equation}
where $\mathcal{L}_{\mathrm{prior}}$ is $\mathcal{L}_{\mathrm{prior}}^{G}$ or
$\mathcal{L}_{\mathrm{prior}}^{PT}$ according to the configuration, with latent
consistency applied as a separate encoder-only update,
\begin{equation}
\mathcal{L}_{E,\mathrm{lat}} = \lambda_{\mathrm{lat}} \mathcal{L}_{\mathrm{lat}} .
\end{equation}

\subsubsection{Base Anomaly Score}
\label{sec:base-score}

After training, anomaly detection uses the deterministic analysis pathway alone.
For an input $x$, the encoder produces $W_0 = E(x)$ and the generator its
reconstruction $\hat{x}_0 = G(W_0)$. The base anomaly score combines
reconstruction error with the discriminator's assessment of realism,
\begin{equation}
s_0(x) = \alpha \left\| x - \hat{x}_0 \right\|_1 - \beta\, D_{\mathrm{logit}}(x) ,
\label{eq:base-score}
\end{equation}
so that higher scores indicate greater deviation from the learned benign
manifold: inputs that reconstruct poorly, that the discriminator judges
unrealistic, or both. This score is the unrefined baseline for all three
configurations, and the starting point for the refinement of
Section~\ref{sec:refinement}.

\subsection{Post-Training Latent Refinement}
\label{sec:refinement}

Refinement serves a diagnostic purpose: it measures how much work is needed to
bring an input's encoded style toward a more representative region of the benign
manifold, once the detector is fixed. Benign and anomalous
inputs are expected to differ in this effort. A benign input is more likely to already lie near an appropriate
benign representation and needs little adjustment, whereas an anomalous input must
be moved further before its reconstruction improves. The number of refinement
steps required, and the distance travelled, therefore carry evidence of
abnormality, which we fold into the anomaly score
(Section~\ref{sec:refined-score}). 

Refinement is applied only after a SAGEGAN
configuration has been fully trained, and it changes nothing about the detector
itself: the encoder, generator, discriminator, style sampler, and EMA generator
are all frozen. The only quantity it alters is the encoded style tensor assigned
to a given input. Related latent optimisation has been used in GAN inversion and
generative anomaly detection \cite{Schlegl2017AnoGAN,Creswell2018GANInversion}.
Here, it operates directly on the layer-wise SAGEGAN style tensor produced by the
encoder's analysis pathway (Section~\ref{sec:encoder}).

We consider two refinement methods: a deterministic greedy search and a learned
policy. These methods draw on the same three ingredients: a fixed geometry of the
benign style space, a state describing the current position within it, and a
discrete set of actions for moving through it. We describe these shared
ingredients first, then the two methods that use them, and finally the anomaly
score that incorporates the outcome.

\subsubsection{Benign Reference Geometry}
\label{sec:geometry}

All statistics defining the refinement space are estimated from encoded benign
\emph{training} styles only, so that refinement, like the detector it operates on,
never sees malware labels. Let
\begin{equation}
\mathcal{W}_{B} = \{E(x) : x \in \mathcal{B}_{\mathrm{train}}\}
\end{equation}
denote the deterministic benign training styles. For each style layer $l$ we
estimate a coordinate-wise mean $\mu_l$ and standard deviation $\sigma_l$. The
normalised distance of a current style tensor $W_t$ from the benign reference at
layer $l$ is
\begin{equation}
d_l(W_t) = \sqrt{\frac{1}{d}\sum_{j=1}^{d}
\left(\frac{w_{t,l,j}-\mu_{l,j}}{\sigma_{l,j}+\epsilon}\right)^2},
\end{equation}
and these are aggregated across layers into a single prior-distance term,
\begin{equation}
d_{\mathrm{prior}}(W_t) = \frac{1}{7}\sum_{l=1}^{7} d_l(W_t)^2 .
\end{equation}
Principal component analysis is additionally fitted to the flattened benign
training styles. Writing $v_1,\ldots,v_q$ for the retained principal directions
and $\mu_{\mathrm{flat}}$ for the mean flattened style, the PCA coordinates of
$W_t$ are
\begin{equation}
c_k(W_t) = \left(\mathrm{vec}(W_t) - \mu_{\mathrm{flat}}\right)^{\top} v_k .
\end{equation}
This construction applies unchanged to all three configurations, its
dimensionality following the trained model automatically: $7\times512$ for $G512$
and $7\times48$ for $G48$ and $PT48$.

\subsubsection{Refinement State and Action Space}
\label{sec:state-actions}

At refinement step $t$, the state summarises reconstruction quality, proximity to
the benign manifold, and position within the refinement trajectory,
\begin{equation}
\begin{aligned}
s_t = \bigl[&d_1(W_t),\ldots,d_7(W_t),\\
&e_t,\;D_{\mathrm{logit}}(x),\;d_{\mathrm{prior}}(W_t),\\
&t/T,\;c_1(W_t),\ldots,c_q(W_t)\bigr],
\end{aligned}
\end{equation}
where $e_t = \|x - G(W_t)\|_1$ is the current reconstruction error and $T$ is the
maximum number of steps. The discriminator response $D_{\mathrm{logit}}(x)$ is
computed once from the original input and held fixed, so that refinement cannot
manipulate the discriminator directly and is confined to changes in the encoded
style.

The action space is built from the same benign geometry, and comprises three
kinds of move. A no-op leaves the style unchanged. For each layer $l$, two actions
step the style toward or away from that layer's benign mean, along
\begin{equation}
u_l^{+} = \frac{\mu_l - w_{t,l}}{\|\mu_l - w_{t,l}\|_2 + \epsilon},
\qquad u_l^{-} = -\,u_l^{+} .
\end{equation}
For each retained PCA direction, two further actions step along $\pm v_k$. The
total number of discrete actions is thus
\begin{equation}
|\mathcal{A}| = 1 + 2L + 2q, \qquad L = 7 ,
\end{equation}
each non-zero action applying a fixed step of magnitude $\eta$. Because every
direction is derived from benign training styles, the action set requires no
malware labels to define movement through the latent space.

\subsubsection{Greedy Refinement}
\label{sec:greedy}

The greedy method is a non-learned control: it establishes how much of any
refinement benefit follows simply from the availability of a useful local search
space, before any policy is learned. At each step it evaluates every candidate
action and scores the resulting style tensor $W'$ by
\begin{equation}
\begin{aligned}
J_{\mathrm{greedy}}(W') &= e(W')\\
&\quad + \lambda_p\,d_{\mathrm{prior}}(W')\\
&\quad + \lambda_s\,\|W'-W_t\|_{\mathrm{RMS}},
\end{aligned}
\end{equation}
which balances reconstruction improvement against distance from the benign
reference and the size of the move. The action minimising $J_{\mathrm{greedy}}$ is
taken only if it improves on the current style; otherwise the no-op is retained.
This repeats for $T$ steps. As nothing is trained, any improvement it produces
measures local latent navigability alone.

\subsubsection{Reinforcement-Learning Refinement}
\label{sec:rl}

The learned method replaces greedy enumeration with a policy over the same action
space, to test whether a learned strategy improves on local search. We use
REINFORCE, a Monte Carlo policy-gradient method \cite{Williams1992REINFORCE}. A
policy $\pi_{\theta}(a_t \mid s_t)$, represented by a multilayer perceptron,
maps the state to logits over the discrete actions; during training an action is
sampled from the resulting categorical distribution. The immediate reward rewards
reconstruction improvement while penalising unnecessary movement away from the
benign geometry,
\begin{equation}
\begin{aligned}
r_t &= (e_t-e_{t+1})\\
&\quad - \beta_s\|W_{t+1}-W_t\|_{\mathrm{RMS}}\\
&\quad - \beta_p\,d_{\mathrm{prior}}(W_{t+1})\\
&\quad - \beta_a\,\mathbb{I}[a_t\neq a_{\mathrm{noop}}],
\end{aligned}
\end{equation}
where $\beta_s$, $\beta_p$, and $\beta_a$ weight latent displacement, post-action
distance from the benign reference, and the use of a non-zero action,
respectively. With a moving baseline $b$ to reduce gradient variance, the training
objective is
\begin{equation}
\mathcal{L}_{RL} = -\,\mathbb{E}\!\left[\log \pi_{\theta}(a_t \mid s_t)\,(r_t - b)\right].
\end{equation}
As with the detector, the policy is trained on benign samples only; malware
validation examples inform model selection and score calibration but never update
the policy, preserving the benign-only learning assumption. At inference the
policy is made deterministic by selecting the highest-logit action at each step,
\begin{equation}
a_t^{*} = \arg\max_a \pi_{\theta}(a \mid s_t).
\end{equation}
Because greedy and RL share the action space, the step size, and the number of
steps, any performance difference between them is attributable to the policy
rather than to the search space available to it.

\subsubsection{Refinement-Aware Anomaly Scoring}
\label{sec:refined-score}

Refinement yields information beyond the base score of
Section~\ref{sec:base-score}: not only the reconstruction finally achieved, but
how much work was needed to achieve it. We retain the total latent displacement
\begin{equation}
\Delta_W = \|W_T - W_0\|_{\mathrm{RMS}},
\end{equation}
the number of non-zero actions applied $n_{\mathrm{act}}$, the final prior
distance $d_{\mathrm{prior}}(W_T)$, and the final reconstruction error $e_T$, and
combine them with the base score into
\begin{equation}
s_R(x) = s_0(x)
+ \gamma_{\Delta}\Delta_W
+ \gamma_p\, d_{\mathrm{prior}}(W_T)
+ \gamma_n\, n_{\mathrm{act}}
+ \gamma_r\, e_T .
\end{equation}
The coefficients $\gamma_{\Delta}$, $\gamma_p$, $\gamma_n$, $\gamma_r$ are selected
on validation data alone and frozen before test evaluation, so that the test set
never influences how the signals are combined.

The displacement and action-count terms are what distinguish this score from a
reconstruction-only criterion. An input already lying near an appropriate benign
representation is reconciled with the manifold in few actions and little movement;
one that must be moved a long way, over many actions, before its reconstruction
improves provides different evidence, even if the two reach a similar final
reconstruction. By recording how readily each input can be brought to the
benign manifold,
refinement contributes evidence of abnormality that the base score does not capture.

\section{Experimental Design}
\label{sec:experimental_design}

\subsection{Dataset and Repeated-Run Protocol}

We used the primary Windows portable executable (PE) corpus described in
SAGEGAN~\cite{Wijayasiri2026SAGEGAN}. The corpus contains 20,000 benign
executables from the Michael Lester collection and 10,820 malicious
executables from MalwareBazaar spanning 214 malware families. Both
groups were selected from the 2--50~MB file-size range. We used the
precomputed SAGEGAN image representation.

Each executable was represented by a three-channel image containing the raw-byte, benign bigram-surprise, and entropy-deviation channels described in that work. Images were loaded as RGB, resized to $256\times256$ pixels, and normalized to $[-1,1]$ in each channel.

The evaluation protocol partitioned benign images into 50\% training, 25\% validation, and 25\% test subsets. Malware images were divided into 10\% validation and 90\% test subsets. Only benign training images were used to update detector and refinement-policy parameters. Both classes were present in validation for checkpoint selection, score calibration, and threshold selection. Thus, parameter learning was benign-only, while model selection used labelled validation data. Test labels were used only for final evaluation.

Experiments used ten seeds. Data partitions varied across seeds, so between-run variability reflects both data partitioning and stochastic training. Each model variant was trained once per seed, giving 30 base-model runs.

Training augmentation consisted of horizontal flipping with probability 0.5 and colour jitter with brightness, contrast, and saturation strengths of 0.05 and no hue jitter. Validation, testing, latent diagnostics, and refinement-policy training used deterministic resizing and normalization without augmentation.

\subsection{Base-Model Training and Checkpoint Selection}

All configurations used the SAGEGAN encoder, generator, and discriminator described in Section~\ref{sec:methodology}. Table~\ref{tab:experimental_settings} gives the common optimization settings. Each minibatch involved four updates: discriminator training with real and generated images; adversarial generator and sampler training; reconstruction and prior-alignment training; and an encoder-only latent-consistency update. Reconstruction used mean absolute error, prior alignment used coordinate-averaged squared moment differences, and latent consistency used mean squared error. Generated images and target styles were detached during the encoder-only update.

\begin{table}[t]
\centering
\caption{Common base-model and REINFORCE training settings. Sampler parameters include the learned affine transformations and, for Gaussian models, the base distribution parameters.}
\label{tab:experimental_settings}
\small
\begin{tabular}{@{}p{0.57\linewidth}p{0.36\linewidth}@{}}
\toprule
Setting & Value \\
\midrule
Base training epochs / batch size & 500 / 8 \\
Encoder, generator, discriminator learning rate & $10^{-4}$ \\
Sampler learning rate & $2.5\times10^{-5}$ \\
Base Adam coefficients $(\beta_1,\beta_2)$ & $(0,0.99)$ \\
Reconstruction / prior / consistency weights & $10\;/\;0.01\;/\;1$ \\
R1 coefficient / application interval & $10$ / 16 minibatches \\
Generator EMA decay & $0.999$ \\
Base validation interval & 5 epochs \\
\midrule
Policy epochs / maximum minibatches per epoch & 20 / 200 \\
Policy batch size / learning rate & 16 / $10^{-4}$ \\
Policy Adam coefficients $(\beta_1,\beta_2)$ & $(0.9,0.999)$ \\
Policy hidden layers & $256,256$ \\
Policy gradient norm limit & 5 \\
Policy validation interval & 5 epochs \\
Refinement steps / non-zero action step size & 3 / $0.05$ \\
\bottomrule
\end{tabular}
\end{table}

R1 regularization contributed $\frac{10}{2}\|\nabla_x D(x)\|_2^2$, averaged over benign images, on every sixteenth minibatch, without an additional interval multiplier. An exponential moving average (EMA) of generator parameters was updated after every minibatch and used for evaluation. Training used automatic mixed precision, fixed learning rates, and the full 500-epoch schedule. The checkpoint with the highest validation ROC AUC, evaluated every five epochs, was retained; ties retained the earlier checkpoint.

The base anomaly score combined mean reconstruction error and the negative discriminator logit with unit weights. At the selected checkpoint, the classification threshold maximized validation Youden's $J=\mathrm{TPR}-\mathrm{FPR}$. The selected model and threshold were then frozen for test evaluation.

\subsection{Refinement and Validation-Based Score Calibration}

Greedy and REINFORCE refinement each started from the same selected base checkpoint for a given variant and seed. Reference means, standard deviations, and 16 principal-component directions were estimated from encoded benign training images only. Principal components were computed from centred, flattened $7d$-dimensional style tensors. Both methods used the same 47 actions: a no-op, movement towards or away from the benign mean in each of seven layers, and movement in either direction along each of 16 principal components. Non-zero actions had Euclidean step size 0.05, and validation and test refinement applied three successive actions.

Greedy refinement evaluated all candidate actions at each step, selecting an improvement in reconstruction error plus benign-reference and displacement penalties, each weighted by 0.1. REINFORCE used a 27-dimensional state and a categorical policy with two 256-unit hidden layers, layer normalization, and leaky-ReLU activations with slope 0.2. Its immediate reward was the reconstruction improvement minus penalties of 0.1 times the RMS step displacement, 0.1 times the post-action benign-reference distance, and 0.005 for a non-zero action. The reward baseline was initialized to zero and updated as $b\leftarrow0.95b+0.05\overline{r}$ for each minibatch.

Policy training sampled one action from each image's initial encoded state and optimized its immediate reward using the REINFORCE objective. It did not train on three-step returns. During validation and testing, the learned policy was applied sequentially for three steps, selecting the highest-logit action each time. The methods therefore shared the action space and number of refinement steps, but greedy enumeration and policy inference had different computational costs.

For each method, variant, and seed, the refinement-aware score coefficients defined in Section~\ref{sec:methodology} were selected by exhaustive validation search over
\[
\gamma_{\Delta}\in\{0,0.5,1,2,5\},\qquad
\gamma_p\in\{0,0.1,0.5,1,2\},
\]
\[
\gamma_n\in\{0,0.01,0.05\},\qquad
\gamma_r\in\{0,0.25,0.5\}.
\]
These grids gave 225 combinations. Selection maximized validation AUC, using validation balanced accuracy to break ties. For REINFORCE, this calibration was repeated at each policy validation checkpoint, and the policy with the highest resulting validation AUC was retained. The final score weights and validation-selected Youden threshold were frozen before testing. Reported refinement gains therefore measure the complete refinement-and-calibration procedure, including the additional score components.

\subsection{Latent Diagnostics and Statistical Analysis}

Four latent diagnostics were recorded after every base-training epoch using the statistics of the encoded data. Encoded statistics used the first four unshuffled benign training batches, giving 32 images with deterministic transforms. The same image subset was retained throughout each diagnostic run. Their seven layer vectors were pooled into a $224\times d$ matrix. 

\paragraph{ROC AUC}
The primary detection metric was test ROC AUC. AUC and refinement gains were summarized as mean $\pm$ sample SD across the ten seeds, with each gain calculated relative to the same run's base AUC before aggregation. Paired comparisons focus on $PT48$ versus the two Gaussian controls ($G48$ and $G512$) using within-seed differences, two-sided paired $t$-tests, and unadjusted 95\% confidence intervals. These summaries characterize variation across repeated runs on the available corpus.

\paragraph{Effective rank}
After centring each coordinate, effective rank was computed from singular-value entropy, $r_{\mathrm{eff}}=\exp(-\sum_j p_j\log p_j)$, where $p_j=s_j/\sum_k s_k$. The reported normalized rank divided this value by $\min(224,d)$: 224 for $G512$ and 48 for $G48$ and $PT48$. It describes the pooled layer-vector matrix rather than the rank of a flattened $7d$-dimensional representation.

\paragraph{Prior mean and variance RMSE}
Prior mean and variance RMSE compared the pooled encoded moments with moments from 256 sampled style tensors, using population variances. The RMSE was calculated across the $d$ coordinates. These diagnostics used sampled, affine-transformed styles as the reference for all three variants.

\paragraph{Latent consistency}
Latent consistency measured whether a style vector can be recovered after it is used to generate an image. We sampled 64 style tensors, generated images with the EMA generator, and re-encoded those images to obtain recovered styles. This diagnostic reports the mean squared error between the recovered and sampled styles, averaged over layers and coordinates. 

\paragraph{}
Diagnostic curves report the mean and one standard deviation (SD) across ten seeds at each epoch. Diagnostic sampling used the same random seed as was used for training. For late-training summaries, epochs 401--500 were first averaged within each seed and then aggregated across seeds. Summaries over epochs 1--500 and 251--500 provided additional checks of the observed ordering. Epochs were not treated as independent experimental replicates, and these windows were separate from validation-selected detection checkpoints.

\subsection{Implementation and Hardware}

Classical model training, refinement, and evaluation were performed using an NVIDIA RTX A6000 GPU, an Intel Xeon w9-3495X CPU, and 256\,GB of RAM. The implementation used PyTorch and torchvision, with NumPy and scikit-learn for numerical processing and evaluation. Python, NumPy, and PyTorch random-number generators were seeded for each run. cuDNN benchmarking was enabled and deterministic execution was not enforced. 

We acquired the photonic samples remotely using an ORCA Computing PT-2,
a rack-mounted photonic quantum system designed for room-temperature
deployment. We used its time-bin interferometer with
48 temporal modes (qumodes) and a $(1,2)$ loop configuration.

\section{AUC Results}
\label{sec:results}

We compare the test area under the receiver operating characteristic curve (AUC) of the three latent configurations before and after post-training refinement. Table~\ref{tab:aggregate_test_auc} reports the mean and sample standard deviation (SD) across the same ten seeds for the base, greedy, and reinforcement learning (RL) conditions. The SD describes variation between runs. Refinement gains are calculated relative to each seed's corresponding base model, with their mean and SD then calculated across the ten paired gains.

\begin{table*}[t]
\centering
\caption{Test AUC across ten seeds, reported as mean $\pm$ sample SD. AUC and gains are expressed on the 0--1 scale. Gain columns report the mean and SD of paired improvements over the corresponding unrefined base model. Bold values indicate the highest mean in each column; statistical comparisons are reported separately. The RL condition uses REINFORCE.}
\label{tab:aggregate_test_auc}
\resizebox{\linewidth}{!}{%
\begin{tabular}{lccccc}
\toprule
Model & Base AUC & Greedy AUC & Greedy gain & RL AUC & RL gain \\
\midrule
$G512$ &
$0.8420 \pm 0.0177$ &
$0.8702 \pm 0.0177$ &
$0.0282 \pm 0.0030$ &
$0.8713 \pm 0.0178$ &
$0.0293 \pm 0.0036$ \\

$G48$ &
$\mathbf{0.8586} \pm 0.0154$ &
$0.8885 \pm 0.0173$ &
$0.0298 \pm 0.0045$ &
$0.8905 \pm 0.0167$ &
$0.0318 \pm 0.0048$ \\

$PT48$ &
$0.8570 \pm 0.0136$ &
$\mathbf{0.8922} \pm 0.0146$ &
$\mathbf{0.0352} \pm 0.0061$ &
$\mathbf{0.8935} \pm 0.0147$ &
$\mathbf{0.0365} \pm 0.0061$ \\
\bottomrule
\end{tabular}%
}
\end{table*}

\subsection{Base Detection Performance}

Before refinement, $G48$ and $PT48$ achieve similar AUCs and both outperform the original Gaussian configuration $G512$. These results indicate that a smaller-sized prior improves results, irrespective of the nature of the prior distribution.

\subsection{Performance After Refinement}

Refinement increases the mean test AUC for all three configurations. After both greedy and RL refinement, the highest mean AUC is obtained by $PT48$, followed by $G48$ and $G512$.

The largest mean improvements over the base model are observed for $PT48$, with gains of 0.0352 for greedy refinement and 0.0365 for RL refinement. The corresponding gains are 0.0298 and 0.0318 for $G48$, and 0.0282 and 0.0293 for $G512$. These values evaluate latent refinement together with the associated scoring and calibration procedure described in Section~\ref{sec:methodology}. The additional mean AUC obtained with RL over greedy refinement is modest, suggesting that greedy refinement therefore accounts for most of the mean improvement observed under the RL condition.

We also observe that, across seeds, $PT48$ has the lowest observed SD of test AUC in each inference condition. After RL refinement, its SD is 0.0147, compared with 0.0167 for $G48$ and 0.0178 for $G512$.

\subsection{Statistical Analysis}

\begin{table*}[htbp]
\centering
\caption{Statistical comparisons of base and RL-refined test AUC across the same ten seeds, focusing on $PT48$ versus the two classical Gaussian controls. Positive differences favour $PT48$. Confidence intervals and $p$-values are unadjusted.}
\label{tab:prior_comparisons}
\resizebox{\textwidth}{!}{%
\begin{tabular}{llrrrl}
\toprule
Comparison & Stage & Mean difference & 95\% CI & $p$ & Seeds A$>$B \\
\midrule
$PT48-G48$ & Base & -0.0017 & $[-0.0180,\ 0.0147]$ & 0.8234 & 5/10 \\
$PT48-G512$ & Base & 0.0150 & $[0.0024,\ 0.0276]$ & 0.0249 & 7/10 \\
$PT48-G48$ & RL final & 0.0030 & $[-0.0136,\ 0.0196]$ & 0.6916 & 5/10 \\
$PT48-G512$ & RL final & 0.0222 & $[0.0083,\ 0.0361]$ & 0.0056 & 9/10 \\
\bottomrule
\end{tabular}%
}
\end{table*}

\begin{table*}[htbp]
\centering
\caption{Statistical comparisons of refinement responsiveness across ten seeds, focusing on $PT48$ versus the two classical Gaussian controls. Gain comparisons contrast the improvement over each model's own base AUC. Positive differences favour $PT48$. Confidence intervals and $p$-values are unadjusted.}
\label{tab:refinement_comparisons}
\resizebox{\textwidth}{!}{%
\begin{tabular}{llrrrl}
\toprule
Comparison & Quantity & Mean difference & 95\% CI & $p$ & Seeds A$>$B \\
\midrule
$PT48-G48$ & Greedy gain & 0.0054 & $[0.0007,\ 0.0100]$ & 0.0294 & 8/10 \\
$PT48-G512$ & Greedy gain & 0.0070 & $[0.0037,\ 0.0104]$ & 0.0010 & 9/10 \\
$PT48-G48$ & RL gain & 0.0047 & $[0.0001,\ 0.0092]$ & 0.0458 & 7/10 \\
$PT48-G512$ & RL gain & 0.0072 & $[0.0038,\ 0.0106]$ & 0.0010 & 9/10 \\
\bottomrule
\end{tabular}%
}
\end{table*}

The statistical comparisons in Tables~\ref{tab:prior_comparisons} and~\ref{tab:refinement_comparisons} focus on the main question of whether the $PT48$ configuration differs from the two classical Gaussian baselines at matched dimensionality ($G48$) and at the original dimensionality ($G512$). We observe no significant difference in mean AUC between $PT48$ and $G48$. In contrast, the refinement-gain comparisons show significantly larger gains for $PT48$ than for both $G512$ and $G48$(Table~\ref{tab:refinement_comparisons}).

\section{Further Analysis}
\label{sec:analysis}

We examine the learned latent representations alongside the detection results to understand how the three configurations differ. The analysis compares latent dimensionalities, prior configurations at matched dimensionality, and responses to post-training refinement. The comparison between $G48$ and $PT48$ is the main control for assessing differences associated with the prior distribution.

\subsection{Validation Performance During Training}

\begin{figure}[t]
    \centering
    \includegraphics[width=\linewidth]{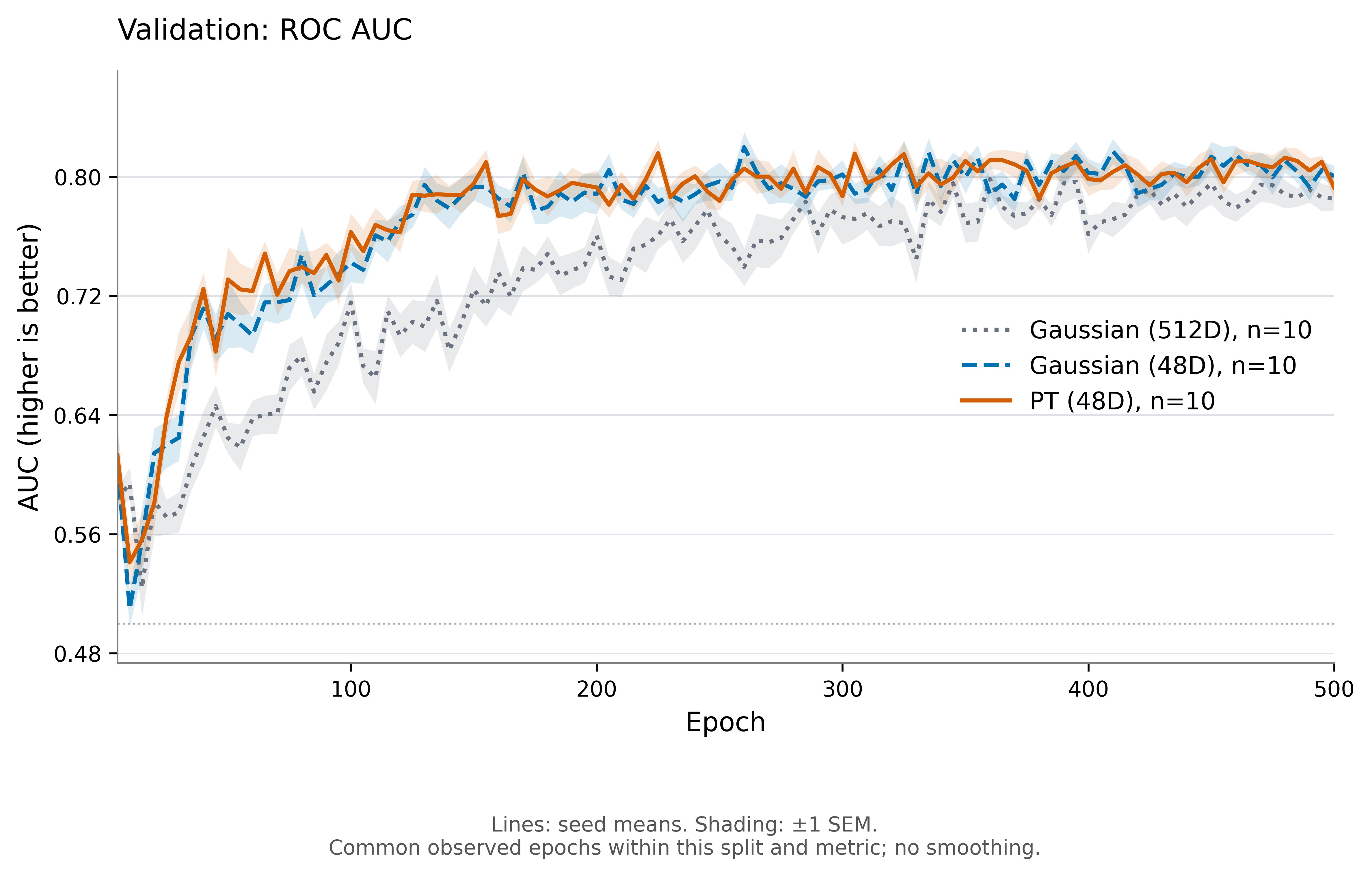}
    \caption{Validation ROC AUC during base-model training
    for $G512$, $G48$, and $PT48$. Lines show the mean
    across ten seeds, and shading represents $\pm$ one
    standard error of the mean (SEM) at each evaluated
    epoch. Curves are shown without smoothing.
    The horizontal dotted line indicates AUC $=0.5$.}
    \label{fig:validation_auc}
\end{figure}

Figure~\ref{fig:validation_auc} shows the mean validation
ROC AUC during base-model training across ten seeds.
Both 48-dimensional configurations achieve higher mean
validation AUC earlier in training than $G512$.
The $PT48$ and $G48$ curves follow similar trajectories
throughout much of training, with smaller improvements
during the later epochs. These observations complement
the final detection results by showing how validation
performance develops before checkpoint selection and
post-training refinement.

\subsection*{Evaluation Metrics}

ROC AUC measures how well the anomaly score ranks malware above benign
files. The following diagnostics characterize the learned styles used
for reconstruction and refinement.

\paragraph{Normalized effective rank.}
Effective rank is computed from the entropy of the normalized singular
values of the centered matrix of pooled benign style vectors. We divide
it by $\min(224,d)$ to obtain normalized effective rank. At a fixed style
width, a higher value indicates that encoded benign variation is spread
more broadly across style directions.

\paragraph{Latent matching.}
For each configuration, we compare the coordinate-wise means and
variances of encoded benign styles with those of sampled styles after
the affine mappings. We report a separate root mean squared error
(RMSE) across coordinates for the means and variances. Lower errors
indicate closer agreement in the location and spread of the styles
supplied to the generator by the two pathways.

\paragraph{Latent consistency.}
We generate an image from a sampled style tensor and re-encode the
image. The mean squared error (MSE) between the original and recovered
styles measures how faithfully the encoder recovers the style used
to generate that image.
\subsection{Latent Geometry and Prior Alignment}

\begin{figure}[htbp]
    \centering
    \includegraphics[width=\linewidth]{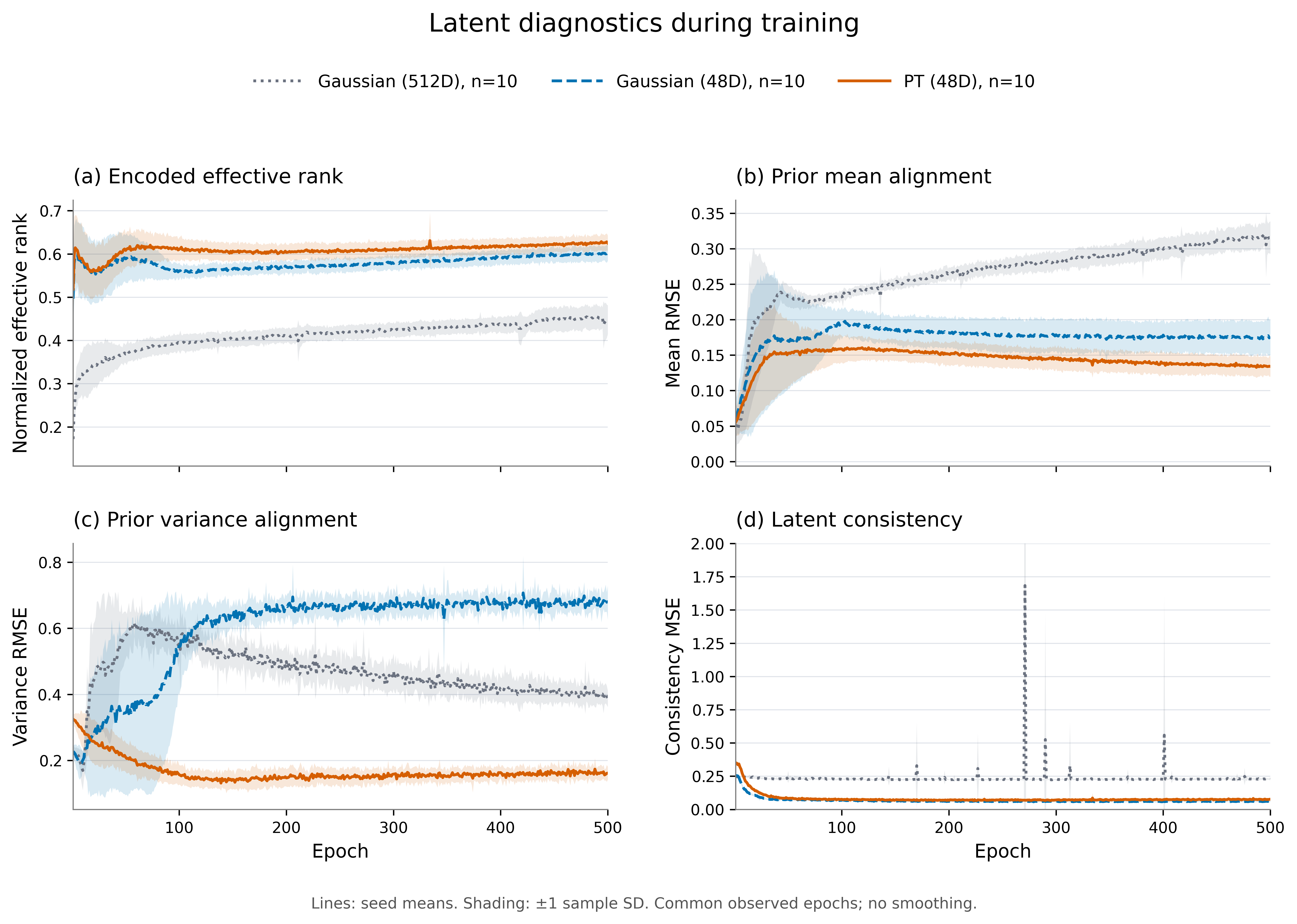}
    \caption{Latent diagnostics during base-model training for $G512$, $G48$, and $PT48$: (a) normalized encoded effective rank, (b) prior mean RMSE, (c) prior variance RMSE, and (d) latent-consistency MSE. Lines show the mean across ten matched seeds and shaded regions show $\pm$ one sample standard deviation at each epoch.}
    \label{fig:latent_diagnostics}
\end{figure}

Figure~\ref{fig:latent_diagnostics} shows four diagnostics over 500 training epochs. Each curve is the mean across the same ten seeds, with shading representing one sample SD across seeds at each epoch. Table~\ref{tab:latent_diagnostics_summary} summarises epochs 401--500. For this table, each diagnostic is first averaged within each seed over the 100 epochs, and the mean and standard error of the mean (SEM) are then calculated across the ten seed averages. Consecutive epochs are therefore not treated as independent experimental runs. These summaries describe the late training period rather than the checkpoints used for the final detection results.

\begin{table*}[htbp]
\centering
\caption{Latent diagnostics over epochs 401--500. Values are the mean $\pm$ SEM across ten seed-level averages. The SEM is calculated across seeds, not across epochs. Rank and error values use the normalization and scale of the logged diagnostics.}
\label{tab:latent_diagnostics_summary}
\resizebox{\textwidth}{!}{%
\begin{tabular}{lcccc}
\toprule
Model & Normalized rank & Mean RMSE & Variance RMSE & Consistency MSE \\
\midrule
$G512$ & $0.4451 \pm 0.0070$ & $0.3094 \pm 0.0051$ & $0.4077 \pm 0.0099$ & $0.2324 \pm 0.0034$ \\
$G48$  & $0.5969 \pm 0.0053$ & $0.1754 \pm 0.0073$ & $0.6770 \pm 0.0075$ & $0.0602 \pm 0.0011$ \\
$PT48$ & $0.6218 \pm 0.0056$ & $0.1362 \pm 0.0046$ & $0.1608 \pm 0.0062$ & $0.0748 \pm 0.0014$ \\
\bottomrule
\end{tabular}%
}
\end{table*}

The $PT48$ model has the highest mean normalized encoded effective rank, reaching 0.6218 over the late training window, compared with 0.5969 for $G48$ and 0.4451 for $G512$. The $PT48$ value exceeds that of $G48$ for all ten seeds when comparing their window averages.

At the same latent width, this increase means that the encoded benign
files exhibit variation across more effective directions. This matters
because the benign encodings define the reference for normal variation
and are used to construct the layer-specific and PCA-based refinement
moves. Effective rank characterizes the pooled layer vectors, while
refinement PCA is fitted to the complete style tensors.

The same ordering is observed when averaging over all 500 epochs and over epochs 251--500. These comparisons concern normalized rank; they do not establish that $PT48$ has a larger absolute effective rank than $G512$.

The PT configuration also shows lower prior mean and variance errors. Its late-window mean RMSE is 0.1362, compared with 0.1754 for $G48$, while its variance RMSE is 0.1608, compared with 0.6770. These correspond to approximately 22.3\% and 76.2\% lower errors, respectively. Both differences favour $PT48$ in all ten paired seeds, and the direction is retained in the longer averaging windows. 

The largest relative improvement is in variance matching, showing
closer agreement in how widely individual style coordinates vary
across the two pathways. The same generator synthesizes images from
sampled styles during adversarial training and reconstructs executables
from encoded styles. The lower errors for $PT48$ quantify a smaller
mismatch in the location and spread of those conditioning inputs.
This matters because reconstruction error contributes to the anomaly
score and guides the refinement procedure.

\subsection{Latent Consistency}

Latent consistency shows a different ordering. The $G48$ model has the lowest late-window MSE at 0.0602, followed by $PT48$ at 0.0748 and $G512$ at 0.2324. The lower error for $G48$ is present in all ten paired seeds when compared with $PT48$. Thus, the higher normalized rank and lower moment errors of the PT representation are not accompanied by the lowest error when recovering styles through generation and re-encoding.

\section{Discussion and Limitations}
\label{sec:discussion}

\subsection{What This Study Shows}

The clearest advantages of $PT48$ at matched dimensionality are in
the learned representation. It distributes benign variation more
broadly across style directions and achieves closer moment agreement
between the encoded and sampled styles used by the generator. Under
the same three-step refinement budget, the mean test AUC gain for
$PT48$ exceeded that of $G48$ by 0.0054 with greedy refinement and
0.0047 with reinforcement learning. These gains measure the complete
refinement and score calibration procedure. The final RL-refined
AUCs remained close at 0.8935 for $PT48$ and 0.8905 for $G48$.

The findings also show why several aspects of the representation must be considered together. Higher effective rank and closer moment alignment do not guarantee lower latent-consistency error, with $G48$ performing better on that diagnostic. Moreover, most of the AUC improvement is obtained with greedy refinement, while RL refinement yields similar final performance once calibration is included. A useful direction is therefore to test which latent properties predict refinement gains and detection performance, and to isolate whether the mechanisms linking latent geometry to refinement responsiveness are causal.

\subsection{Limitations}

A first limitation is that this work is descriptive but does not identify a causal mechanism. We establish empirical differences between latent configurations in multiple diagnostics, but the pathway through which these diagnosed quantities affect refinement behaviour or final AUC are not elucidated. Our results validate the expectation that a quantum latent distribution achieves different result than classical, but the specifics of how the properties of this distribution affect these results are still unclear.

The results also do not establish generalisation beyond the specific experimental setting studied here. All evaluations use one malware corpus, one model architecture, and a fixed PT-2 measurement pool collected under one interferometer configuration. However, these results build evidence that differences in behaviour can often be expected to occur when using a quantum latent distribution, which can affect final outcomes.

A further limitation is that we only compare against a simple Gaussian prior rather than a broader class of classical latent distributions. In principle, one could introduce additional classical latents with richer structure, for example, mixtures, autoregressive priors, or flow-based priors. Doing so would help identify which improvements are attributable to quantum-generated correlations versus which are a consequence of using a more expressive non-Gaussian prior. However, we note that comparisons to other distributions were done in prior work \cite{Bacarreza2026QuantumLatent} and is computationally intensive. Our objective here was to explore a promising alternative to the original SAGEGAN model, not search through a large possible design space.

\subsection{Next Steps in Quantum Hardware}

An important observation is that the number of photons in PT-2 is small enough that the system can still be simulated classically. Our work shows only that the output statistics from such a system are beneficial for this task, and does not claim a quantum advantage that could not be replicated classically. 

However, larger-scale photonic quantum systems have been demonstrated \cite{zhong2020quantum,madsen2022quantum}. Here, we discuss ways in which this approach may benefit from the availability of more powerful quantum hardware. First, the PT samples used in this study were drawn from a fixed ORCA Computing PT-2 interferometer configuration and treated as a static prior computed ahead of training. This design choice is partly due to the PT-2 system's relatively slow speed. Next-generation systems are expected to run significantly faster, making it practical to sample the quantum distribution inside the training loop rather than from a precomputed dataset. Faster sampling would allow the interferometer parameters themselves to be optimised jointly with the encoder and generator, so that the latent distribution is shaped toward the benign manifold rather than fixed in advance. This improvement in trainability opens the door to algorithmic changes that could convert the refinement gains observed here into a larger and more consistent effect.

The PT-2 configuration used here also has limited-range entanglement, which is physically implemented by two short delay lines. Future systems are expected to produce longer-range entanglement across qumodes. Longer-range entanglement was shown in \cite{Bacarreza2026QuantumLatent} to lead to improved generative performance on a chemistry task. With our results indicating that the structural properties of the latent distribution of PT-2 improve both higher effective rank and closer latent matching than the Gaussian configurations, it would be interesting to see whether the longer-range structure could further improve results.

Finally, the specifications of PT-2 limited our study to 48 qumodes, while next-generation systems will be larger. A larger mode count is not obviously advantageous for this task: our 48-dimensional models outperformed the original 512-dimensional configuration, suggesting that simply enlarging the latent is unlikely to help and may hurt. However, a larger potential qumode count offers additional flexibility. A larger available dimensionality decouples the choice of latent size from the hardware, allowing the dimensionality to be tuned to the anomaly detection task while still drawing on quantum-generated correlations, and leaving room to test whether some intermediate dimensionality improves on 48.

\subsection{Applications to Other Problems}

Next steps should evaluate broader applications within malware detection. The same framework could also be extended towards network intrusion detection, industrial fault detection, and visual defect inspection. In each case, the encoder and generator would need to suit the data, while refinement could measure how much an input's representation changes relative to learned normal behaviour.

\section{Conclusion}
\label{sec:conclusion}

This study examined the role of a photonic quantum-derived prior in generative anomaly detection. Building on SAGEGAN, we compared 512-dimensional and 48-dimensional Gaussian configurations with a 48-dimensional prior sampled from an ORCA Computing PT-2 processor. Evaluation across ten seeds considered detection performance, latent diagnostics, and the response to greedy and reinforcement-learning refinement while the trained detector remained frozen.

The PT configuration consistently showed higher normalized effective rank and lower prior mean and variance errors than the dimension-matched Gaussian model under the measured diagnostics. These differences did not extend to every property, with the Gaussian model achieving lower latent-consistency error. The combined refinement and scoring procedure improved mean test AUC for all configurations, with the largest gain observed for PT, increasing from 0.8570 to 0.8935 under RL refinement. However, its final AUC remained close to the 0.8905 achieved by the dimension-matched Gaussian model.

These findings show measurable differences when using a quantum latent in the learned representation and its response to refinement. With the latent playing an important role in generative and anomaly detection models, and the growing availability and maturity of quantum hardware, these results reinforce the notion that these hybrid quantum-classical models warrant future study. Future work should focus on better understanding these differences, and determining when and where these differences can help provide a practical benefit for anomaly detection.
The layer-wise styles also provide a way to investigate individual anomaly scores. Starting from the same encoding, future work could move
one style vector at a time toward its benign training mean and measure the resulting change in the reconstruction term of the anomaly score. Residual maps could show where that change occurs across the byte, bigram-surprise, and entropy-deviation channels.

\bibliographystyle{IEEEtran}
\bibliography{references}
\end{document}